\documentclass[12pt]{iopart}
\usepackage{iopams}
\usepackage{epsfig}   
\usepackage{graphics}
\newcommand{\beq}{\begin{equation}}
\newcommand{\eeq}{\end{equation}}
\newcommand{\bea}{\begin{eqnarray}}
\newcommand{\eea}{\end{eqnarray}} 
\newcommand{\ba}{\begin{array}}
\newcommand{\ea}{\end{array}}

\newcommand{\p}{\partial}

\begin{document}

\title[Expansion history and $f(R)$ modified gravity]{Expansion history and $f(R)$ modified gravity}

\author{Malcolm Fairbairn$^{1}$ and Sara Rydbeck$^2$}

\address{$^1$CERN Theory Division, CH-1211, Geneva 23, Switzerland}

\address{$^2$Department of Physics, Stockholm University,\\
Albanova University Center, S--106 91 Stockholm, Sweden}

\begin{abstract}
We attempt to fit cosmological data using $f(R)$ modified Lagrangians containing inverse powers of the Ricci scalar varied with respect to the metric.  While we can fit the supernova data well, we confirm  the $a\propto t^{1/2}$ behaviour at medium to high redshifts reported elsewhere and argue that the easiest way to show that this class of models are inconsistent with the data is by considering the thickness of the last scattering surface.  For the best fit parameters to the supernova data, the simplest $1/R$ model gives rise to a last scattering surface of thickness $\Delta z\sim 530$, inconsistent with observations.
\end{abstract}

\eads{\mailto{malc@cern.ch}, \mailto{sararyd@physto.se}}




\section{Introduction}
Observations of type 1a supernovae still consistently suggest that the universe is accelerating \cite{astier06,riess07,essence} (see \cite{RFG} for a detailed list of references).  This conclusion seems to be supported by the fact that observations of the CMB tell us that the universe is spatially flat \cite{WMAP3} and that the matter which seems to be responsible for galaxy clustering is not enough to account for this flatness \cite{2df,sdss}.  Furthermore, the measurement of space-time geometry achieved by the detection of the imprint of the waves in the primordial plasma in the galaxy correlation function \cite{eisenstein05} has caused problems for models where the supernova data are explained by the dimming of photons \cite{dust,CKT,domainwalls}.

There have been a number of different approaches trying to explain the mystery of the acceleration of the universe.  One such approach consists of a class of theories obtained when the Einstein-Hilbert Lagrangian is modified by hand and the Ricci scalar $R$ is replaced with some function $f(R)$ \cite{capozziello,carroll}.  These theories are phenomenological in as much as it is not clear what the underlying theory that gives rise to them would be.  The motivation is that inverse powers of the curvature in the Lagrangian will give vacuum solutions which are not Minkowski but which rather lead to late time acceleration.  The simplest theory with inverse powers of the curvature in the Lagrangian is
\begin{equation}
S=\frac{M^2_{Pl}}{2}\int d^4x\sqrt{-g}\left(R-\frac{\mu^4}{R}\right)+\int d^4x\sqrt{-g}{\cal L}_M
\label{overRlag}
\end{equation}
where $\mu$ is a mass scale that must be fitted to the data. Variation of this action with respect to the metric gives new field equations
\beq
\hspace{-1cm}\left(1+\frac{\mu^4}{R^2}\right)R_{\mu\nu}-\frac{1}{2}\left(1-\frac{\mu^4}{R^2}\right)Rg_{\mu\nu}+\mu^4[g_{\mu\nu}\nabla_\alpha\nabla^\alpha-\nabla_\mu\nabla_\nu]R^{-2}=\frac{T_{\mu\nu}^M}{M^2_{Pl}}
\label{overRfield}
\eeq
from which it can be seen that for cosmological solutions there will exist a vacuum solution with $H\sim \mu$.  Such $f(R)$ models can therefore give rise to late time acceleration which could be responsible for the apparent dark energy. 

The simplest models of this nature seem to be at odds with solar system tests of gravity, at best containing a light degree of freedom \cite{chiba} and at worst possessing instabilities \cite{dick,dolgov} but one might imagine that the Lagrangian (\ref{overRlag}) is some effective limit valid on very large scales and that some new physics on short distances could change the theory.  There are also indications that some $f(R)$ theories may be safe in some regions of parameter space \cite{navarro,tegmarkchameleon,li} through a process rather similar to the chameleon mechanism \cite{justin} although that does not appear to be the case for the ones studied here.

It should be noted that variation of the normal Einstein-Hilbert Lagrangian with respect to the metric or
alternatively the Christoffel symbols leads to the same field equations for gravity whereas in this class of modified Lagrangians the same is not true.  Field equations obtained using the latter Palatini approach will yield different cosmologies and solar system constraints \cite{palatini,mota,shinjipala}.  In this work we will restrict ourselves to equations of motion obtained by varying the Lagrangian with respect to the metric.

The solution of the field equations for a cosmological background lead to cosmological equations with higher derivative terms, for example for the field equations (\ref{overRfield}) the $tt$ Friedman equation for a spatially flat universe becomes 
\beq
3H^2-\frac{\mu^4}{12(\dot{H}+2H^2)^3}\left(2H\ddot{H}+15H^2\dot{H}+2\dot{H}^2+6H^4\right)=\frac{\rho_M}{M_{Pl}^2}
\label{newfried}
\eeq 
which means that there are more degrees of freedom in the space of solutions than for Einstein gravity with a cosmological constant or the standard DGP model.  This space of solutions needs to be compared with the data.

If type 1a supernovae are good examples of standard candles as is thought, they can in principle trace out the Hubble diagram in an unambiguous fashion.  There is, however, a need for some caution as there are presently ambiguities in the way that one can analyse the data \cite{SALT,MLCS} and also in which supernovae should be included in samples to be used for cosmology \cite{astier06,riess07,essence}.  In certain situations, these ambiguities can lead to different predictions with regards to which dark energy models are favoured over others \cite{fairbairngoobar,scoop,RFG}.

Despite this, all of the supernova surveys seem to agree on some basic facts, namely that the universe is accelerating and that $\Lambda$CDM, i.e. a universe composed of matter, radiation and a constant energy density which does not change over time, fits the data rather well, and better than many alternatives motivated by specific physical models.

Expanding the surveyed redshift range is of great importance to
investigate the nature of dark energy and to find out in particular whether it is a cosmological
constant or not. Thus, the recent compilation of supernova data
in \cite{riess07}, including 23 SNIa at z $\ge$ 1 from an HST/ACS
is a very interesting data set for model comparisons. Riess et al have included data from various sources and re-fitted the light-curves
with the MLCS2k2 technique. 

We have used the gold set of 182 SNe Ia from Riess et al \cite{riess07}, taking
into account the additional redshift error discussed there in the case of the high redshift supernovae for which the redshifts were determined from broad features in the spectra.

Furthermore we have added to this data set the position of the acoustic peak in the SDSS galaxy survey \cite{eisenstein05} and also the CMB shift parameter \cite{wangmukherjee}.  We direct the reader to \cite{RFG} for more details and discussion on measuring the expansion history with these data sets.

This data leads to a set of angular distances and luminosity distances which need to be explained by any successful dark energy model.  In this work we aim to find out if it is possible to fit this data using the theories of modified gravity introduced above.  We also need to verify that the best fit models do not {\it at the same time} change conditions in the early universe such that the CMB and baryon oscillation data is no longer consistent with the physics of that epoch.

In other words, in order for the models to be acceptable, there has to be a circular self consistency between the theory and the data.  It has been observed that if such $1/R$ models give rise to acceleration at late times, then they also lead to expansion at early times which mimics, in terms of effective equation of state, the expansion during the radiation dominated era \cite{shinjiamen1,shinjiamen2} (independent of whether or not there actually is any radiation in the universe).  One question which needs to be answered therefore is what is the density of this effective radiation component for model parameters which also fit the observed luminosity and angular distances in the data.  Having obtained the value of that effective density, we need to see if it is consistent with the observed properties of the CMB.  If not, then the theory may appear to be able to fit the existing data but the data is no longer consistent with the early universe physics implied by the theory, and the model is ruled out.  Finding out if it is possible to evade this situation, and fit all the data simultaneously is the subject of this paper.

\section{Obtaining solutions for $f(R)$ theories.}

There are a large number of different $f(R)$ theories with inverse powers of $R$ that one might consider but we will look at the two simplest models that can be written down, namely $f(R)=R-\mu^4/R$ and $f(R)=R-\mu^6/R^2$. 

In order to test the $f(R)$ theories, we need $H(z)$, the Hubble expansion rate as a function of redshift.  This is rather difficult to obtain from higher order Friedman equations such as (\ref{newfried}).  We choose to perform a conformal transformation which allows us to treat the problem as one of a scalar field $\sigma$ with a potential $V(\sigma)$ evolving in an FRW universe where the matter is endowed with a non-standard gravitational coupling set by $\sigma$.  

Having found the solutions in the Einstein frame for the scale factor, Hubble parameter and the scalar field, we will transform back to the matter frame \footnote{the frame where the matter is not coupled to gravity via $\sigma$, sometimes called the Jordan frame or more recently the string frame} where we will integrate $H(z)$ and compare it with the data.

For an action of the form $f(R)$ with $\partial f/\partial R>0$, we can perform a conformal transformation to the Einstein frame \cite{maeda}.  The conformal transformation is written
\beq
\tilde{g}_{\mu\nu}=pg_{\mu\nu},\quad \frac{\partial f}{\partial R}\equiv p\equiv \exp\left (\sqrt{2 / 3}\sigma\right)
\eeq
we will continue to use both $p$ and $\sigma$ even in the same equations to keep expressions compact despite the fact that they can be used interchangeably.
The equations can now be written in the more familiar form
\beq
\tilde{R}_{\mu\nu}-\frac{1}{2}\tilde{R}\tilde{g}_{\mu\nu}=(\nabla_\mu\sigma)\nabla_\nu\sigma-\frac{1}{2}\tilde{g}_{\mu\nu}\tilde{g}^{\alpha\beta}(\nabla_\alpha\sigma)\nabla_\beta\sigma-V(\sigma)\tilde{g}_{\mu\nu}+\frac{\tilde{T}_{\mu\nu}}{M_{Pl}^2}
\eeq
where the potential and the energy-momentum tensor in the Einstein frame are given by \cite{maeda}
\bea
V&\equiv&\frac{(sign)}{2\left|\partial f/\partial R \right|}\left(R \frac{\partial f}{\partial R}-f\right)\\
\tilde{T}_{\mu\nu}&\equiv&\frac{T_{\mu\nu}}{p}
\eea
with $(sign)= \left.\frac{\partial f}{\partial R}\right/\left|\frac{\partial f}{\partial R} \right|$. The potential $V$ has mass dimension 2 because we have chosen to work with a dimensionless scalar.
We denote quantities in the Einstein frame with a tilde, for example the time coordinate in the Einstein frame Robertson-Walker metric is $d\tilde{t}=\sqrt{p}dt$ and the scale factor $\tilde{a}(t)=\sqrt{p}a(t)$.  The equations which need to be solved simultaneously are the equation of motion for $\sigma$ and the Friedman equation.  The latter comes from the time-time component of the transformed Einstein equations
\beq
3\tilde{H}^2=\frac{1}{2}\sigma'^2+V(\sigma)+\frac{\tilde{\rho}_M}{M_{Pl}^2}
\label{einsfrei}
\eeq
where a prime denotes differentiation with respect to $\tilde{t}$, the density $\tilde{\rho}_M=\rho_M/p^2$ and $\tilde{H}=\tilde{a}'/\tilde{a}$. The equation of motion can be obtained from the time component of the divergence or from the covariant derivative of the stress energy tensor $\tilde{T}$
\beq
\sigma''+3\tilde{H}\sigma'+\frac{\p V}{\p\sigma}-\frac{(1-3w)}{\sqrt{6}}\frac{\tilde{\rho}_M}{M_{Pl}^2}=0
\label{einseom}
\eeq
Equations (\ref{einsfrei}) and (\ref{einseom}) are the ones that we need to solve to get the evolution of the universe in these models. In order to relate the Hubble parameter $H$ in the matter frame and $\tilde{H}$ in the conformal frame, one must use
\beq
H=\sqrt{p}\left(\tilde{H}-\frac{\sigma'}{\sqrt{6}}\right).
\eeq
To compare these models with the data, we need $H(z)/H_0$ in the matter frame, or rather $\int\frac{dz}{H(z)/H_0}$.  There are more free parameters than in $\Lambda$CDM so we restrict ourselves to the case $k=0$. For a flat universe, our free parameters are $\Omega_M$ as defined in the original matter frame, the mass scale $\mu$ and the values today of $\sigma $ and $\sigma '$. To write these in dimensionless form, let us introduce the time-variable $\tau=\tilde{H_0}\tilde{t}$ and the parameters
\beq
\alpha=\frac{H_0}{\mu}\qquad
\beta=\left.1-\frac{1}{\sqrt{6}}\frac{\p\sigma}{\p\tau}\right|_0
\eeq
where the subscript $0$ denotes the values today at redshift $z=0$. If we also introduce the dimensionless potential $U(\sigma)$ so that
\beq
V(\sigma)=\mu^2 U(\sigma)
\eeq
then the $\tau\tau$ Friedman equation becomes
\beq
\frac{\tilde{H}^2}{\tilde{H}^2_0}=\frac{1}{6}\left(\frac{\p\sigma}{\p\tau}\right)^2+\frac{1}{3}\frac{p_0\beta^2}{\alpha^2}U(p)+\frac{\Omega_M\beta^2}{\sqrt{p_0p}}\left(\frac{\tilde{a_0}}{\tilde{a}}\right)^3
\label{frx}
\eeq
while the equation of the motion for the scalar field is written
\beq
\frac{\p^2\sigma}{\p\tau^2}+3\frac{\tilde{H}}{\tilde{H_0}}\frac{\p\sigma}{\p\tau}+\frac{p_0\beta^2}{\alpha^2}\frac{\p U}{\p\sigma}-\sqrt{\frac{3}{2}}\frac{\Omega_M\beta^2}{\sqrt{p_0p}}\left(\frac{\tilde{a}_0}{\tilde{a}}\right)^3=0.
\label{eomx}
\eeq
By solving (\ref{frx}) and (\ref{eomx}) numerically, we can recover the evolution of the Hubble expansion in the Einstein frame and can convert this back to the matter frame using $\sigma(\tau)$.
One of the parameters can be written in terms of the others using (\ref{frx}) with $z=0$
\beq
\alpha^2=\frac{p_0^2\beta U(p_0)}{3(2p_0-p_0\beta-\Omega_M\beta)}
\eeq
We therefore choose to label the parameter space of solutions of $H(z)/H_0$ with the three variables $\Omega_M$, $\sigma_0$ and $d\sigma/d\tau|_0$.

\section{The data vs. $f(R)=R-\mu^4/R$ and $f(R)=R-\mu^6/R^2$}
\begin{figure}
\begin{center}
\includegraphics[height=10cm,width=14cm]{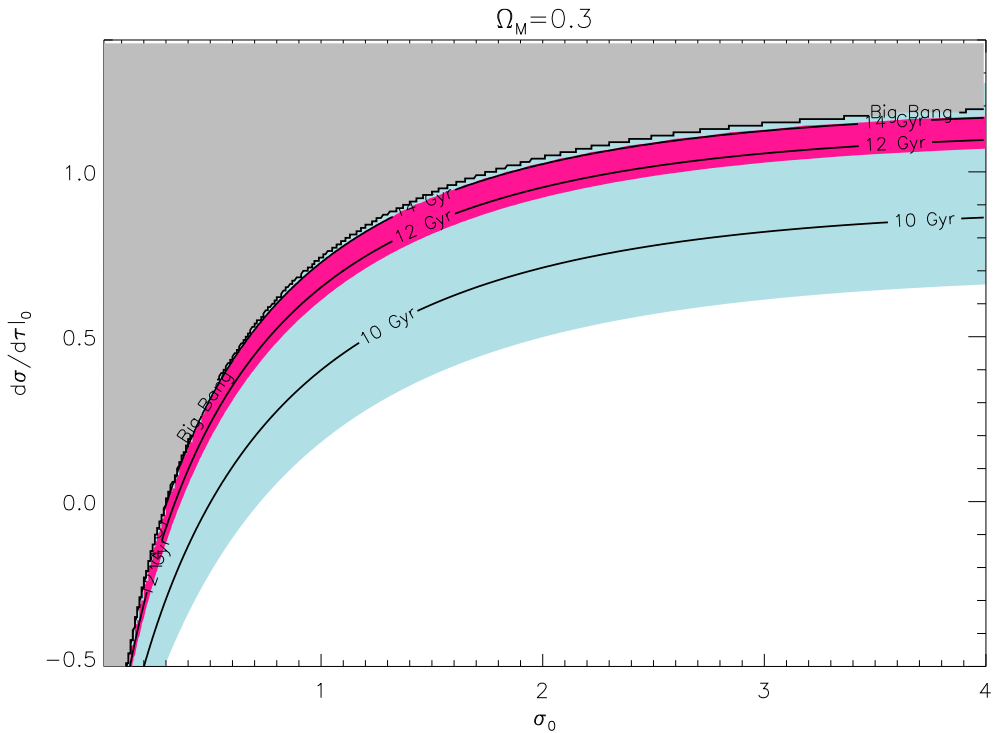} 
\caption{Look back times (age of the universe) for $f(R)=R-\mu^4/R$ cosmologies. We label different cosmologies using the value of the effective scalar $\sigma$ and its time derivative in the Einstein frame $d\sigma/d\tau$ at redshift $z=0$.  These two parameters plus the matter density (this plot is for $\Omega_M=0.3$) provide three of the four parameters necessary to determine the cosmology.  The fourth is the Hubble constant which is integrated over in cosmological fits.  Look-back times are in years in the matter frame assuming a Hubble constant of 65 kms$^{-1}$Mpc$^{-1}$.  In cosmologies above the line labeled 'big bang', the scalar field goes over the top of the potential too quickly (see text).  The red region is the 99\% region fitting to supernova data, the blue region is the same for BAO.}
\label{lookback}
\end{center}
\end{figure}

In this section we will see if it is possible to fit the existing data under the assumption that the effective dark energy density is not significant at early times, in other words that the angular size of the peaks of the CMB and baryon oscillations would be the same as in $\Lambda$CDM.  Later we will go on to explicitly show that this is a bad assumption.

\begin{figure}
\begin{tabular}{cc}
\includegraphics[height=6cm,width=8cm]{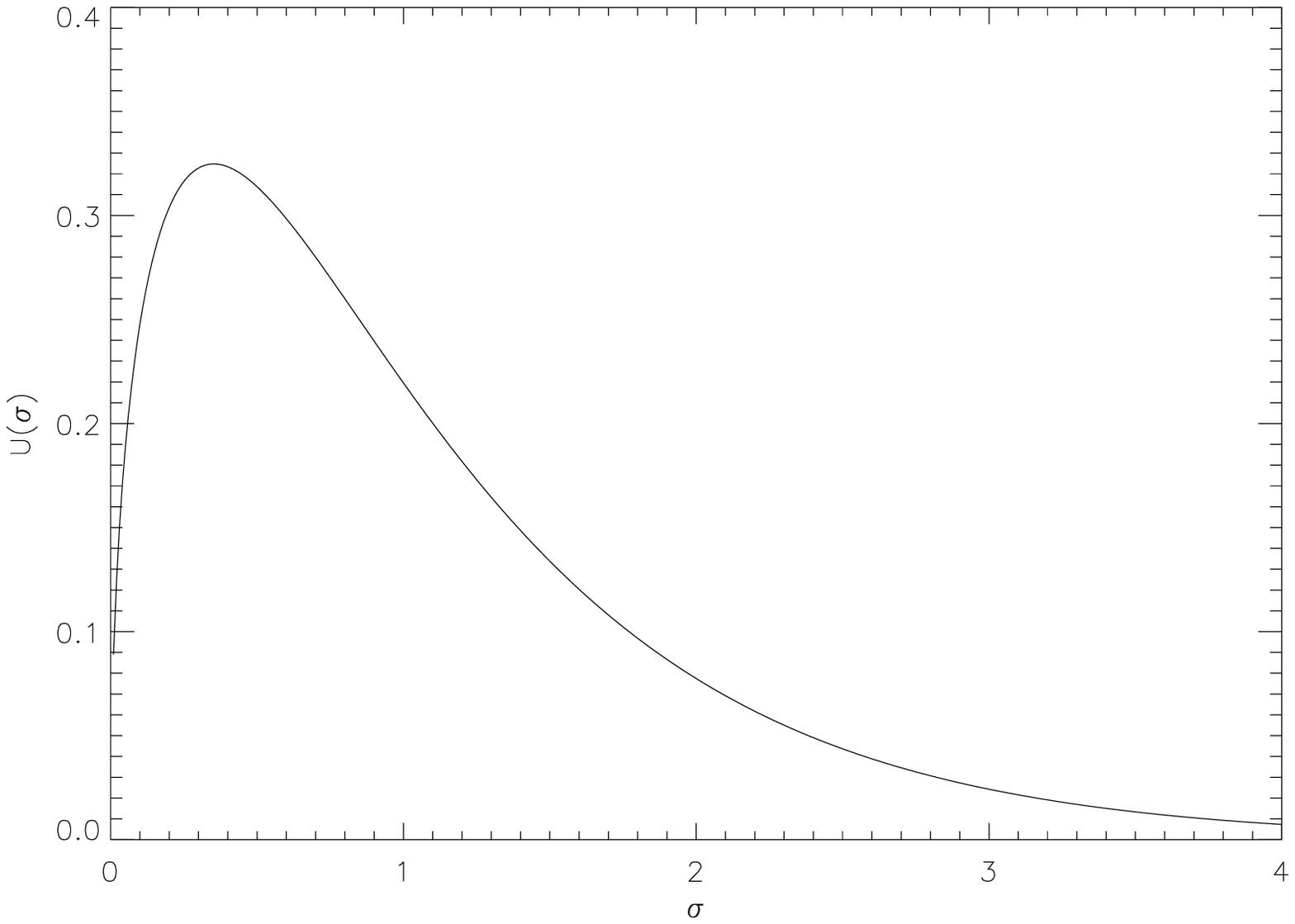}& 
\includegraphics[height=6cm,width=8cm]{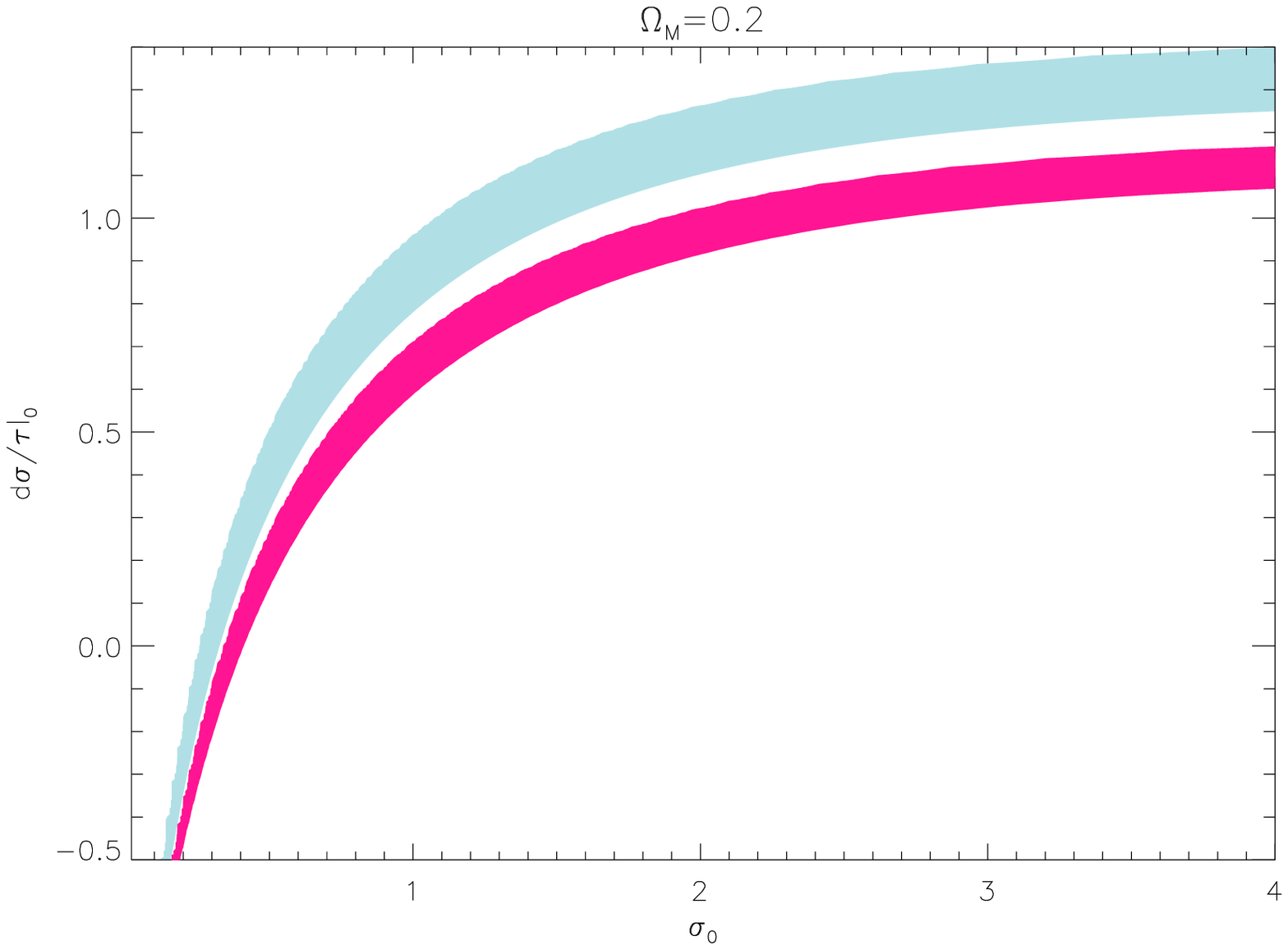}\cr
\includegraphics[height=6cm,width=8cm]{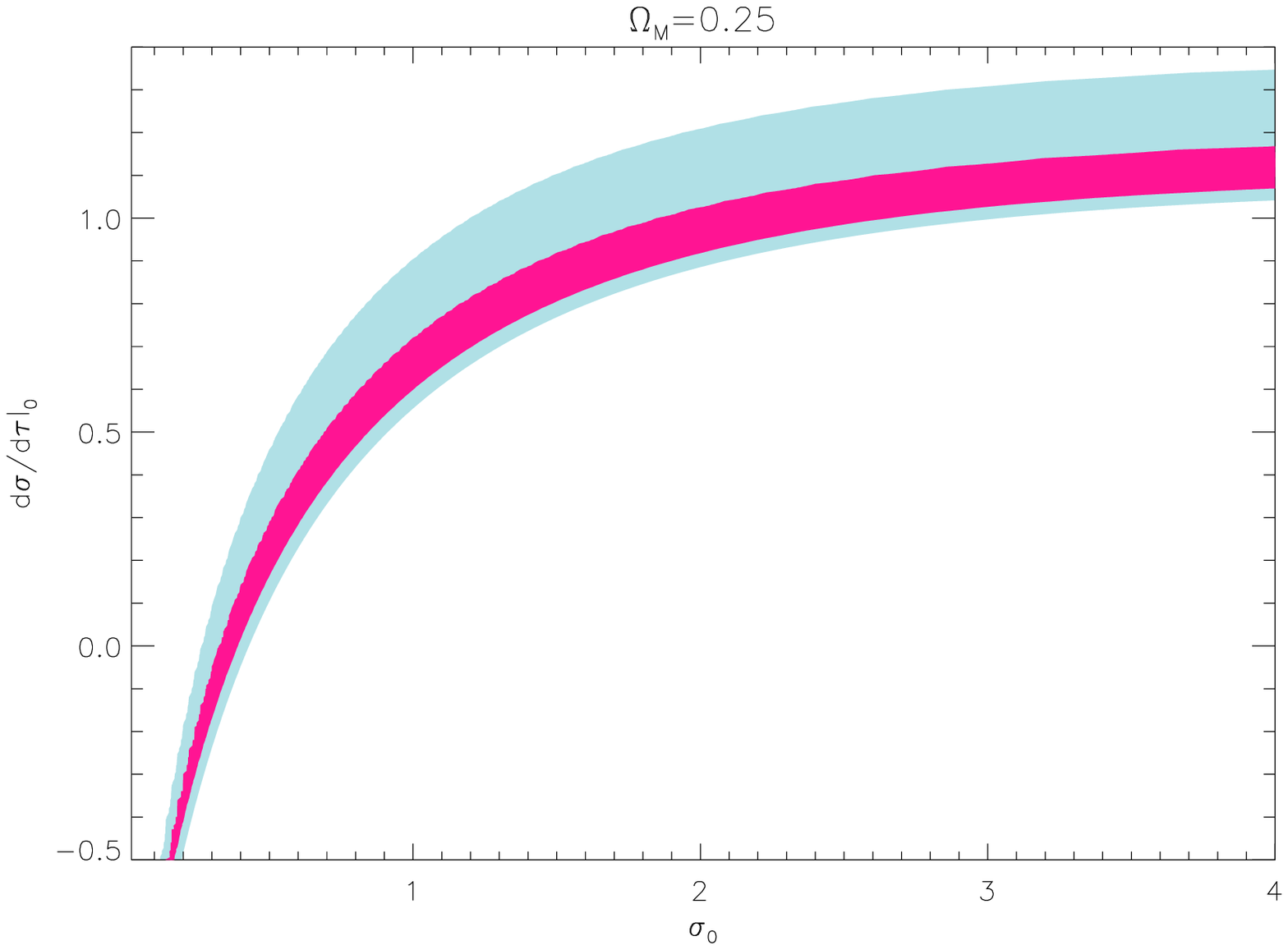}& 
\includegraphics[height=6cm,width=8cm]{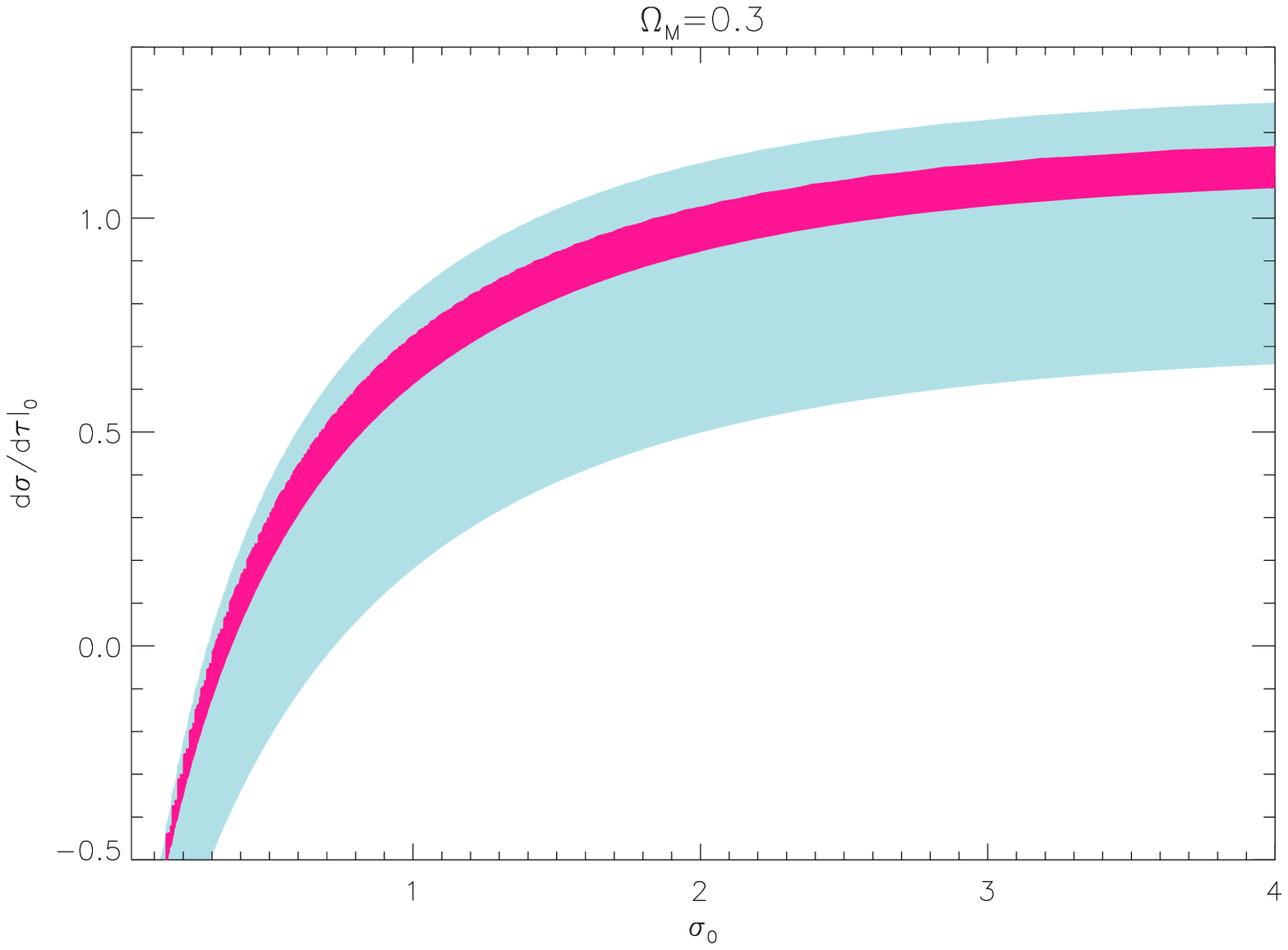}\cr
\end{tabular}    
\caption{Comparison of $f(R)=R-\mu^4/R$ with data.  The top left hand panel is the effective potential in the Einstein frame, while the other three plots are the values of $\sigma_0$ and $d\sigma/d\tau|_0$ favoured by the supernova data in red and the BAO data in blue (see caption of figure \ref{lookback} for more details explaining the plot). The three plots correspond to different matter densities - $\Omega_M=0.2,0.25,0.3$.}
\label{overrres}
\end{figure}

The potential $U(\sigma)$ for the $f(R)=R-\mu^4/R$ model is plotted in the first diagram of figure \ref{overrres}, it rises from zero to a maximum and then falls back to zero as $\sigma$ goes to infinity.

Figure \ref{lookback} shows the combined best fit regions for the supernova, BAO and CMB data for $\Omega_M=0.3$ plotted as a function of $\sigma_0$ and $\left.d\sigma/d\tau\right|_0$.  The best fit to supernova data only is obtained for  $\sigma_0=0.22$ and $\left.d\sigma/d\tau\right|_0=-0.24$, and the best fit when also the two angular distance data points are included for  $\sigma_0=3.20$ and $\left.d\sigma/d\tau\right|_0=1.16$. Plotted also are look back times for the age of the universe in this model assuming a Hubble constant $H_0=65$kms$^{-1}$Mpc$^{-1}$ (or more precisely the elapsed time in the matter frame since a redshift of $z=20$) and a line labeled ``big bang'' which divides the parameter space between the grey region of solutions where the scalar field $\sigma\rightarrow 0$ too quickly in the past, which corresponds to a curvature singularity.  

This can be understood in the following way, for a given value of $\sigma_0$, increasing values of $d\sigma/d\tau$ push $\sigma$ higher up the effective potential as one moves into the past, which means that the rapidly redshifting contribution from $\Omega_M$ can be reduced and the universe expands more slowly in the past, making it older.  This increase in age with $\left.d\sigma/d\tau\right|_0$ ends abruptly for some value of $\left.d\sigma/d\tau\right|_0$ where $\sigma$ goes over the top of the potential and falls to $\sigma=0$ relatively recently in cosmological terms, which signals that $R\rightarrow\infty$ in the matter frame.  That region beyond the curve labeled with the words ``big-bang'' is therefore not included in the analysis and $\chi^2$ values are not calculated there. 

Also plotted are the 99\% confidence bands for the supernova data in red and the baryon oscillation data in blue.  These regions overlap each other for certain values of $\Omega_M$ and therefore seem compatible with the model.

\begin{figure}
\begin{tabular}{cc}
\includegraphics[height=8cm,width=6cm,angle=270]{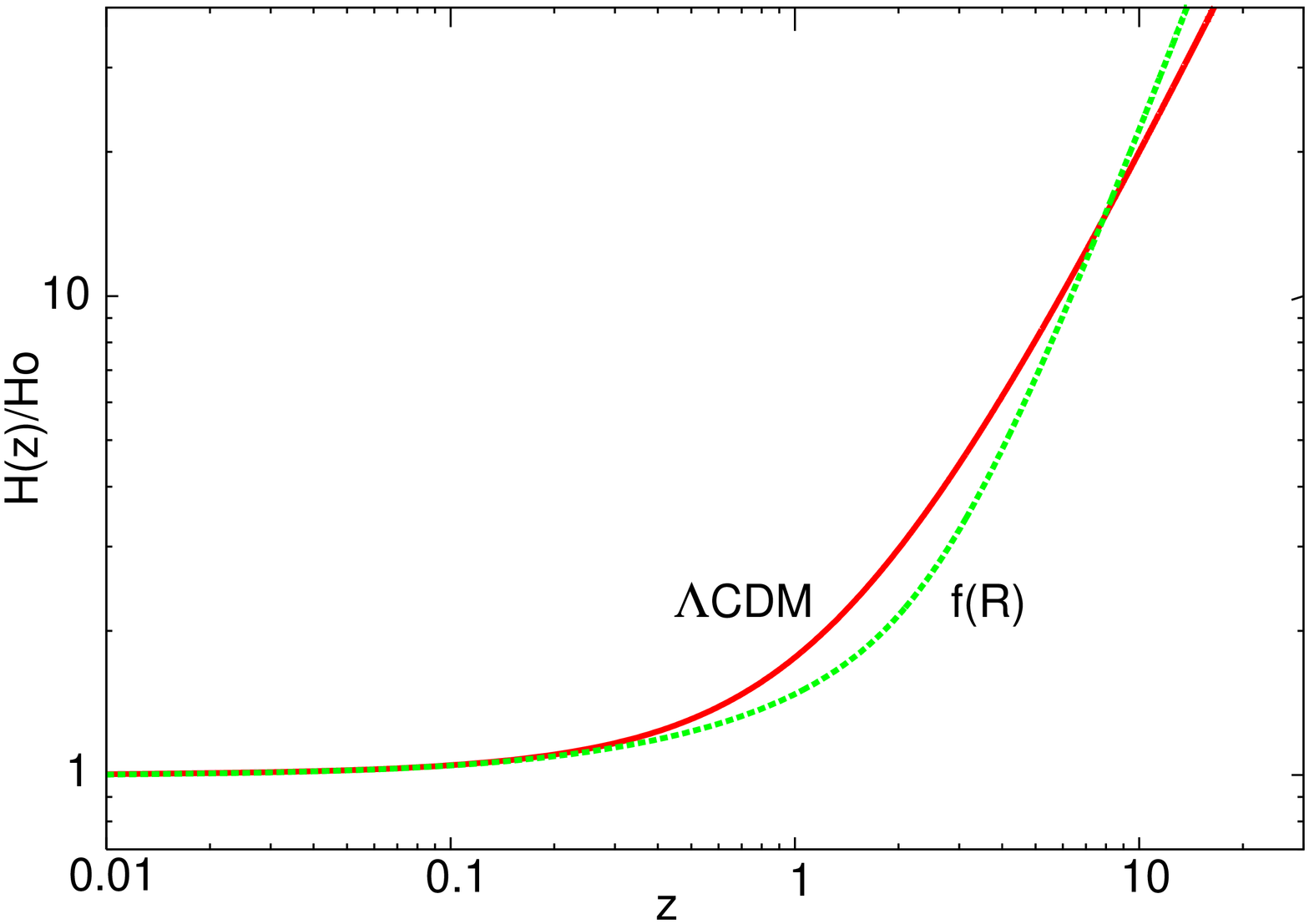}& 
\includegraphics[height=8cm,width=6cm,angle=270]{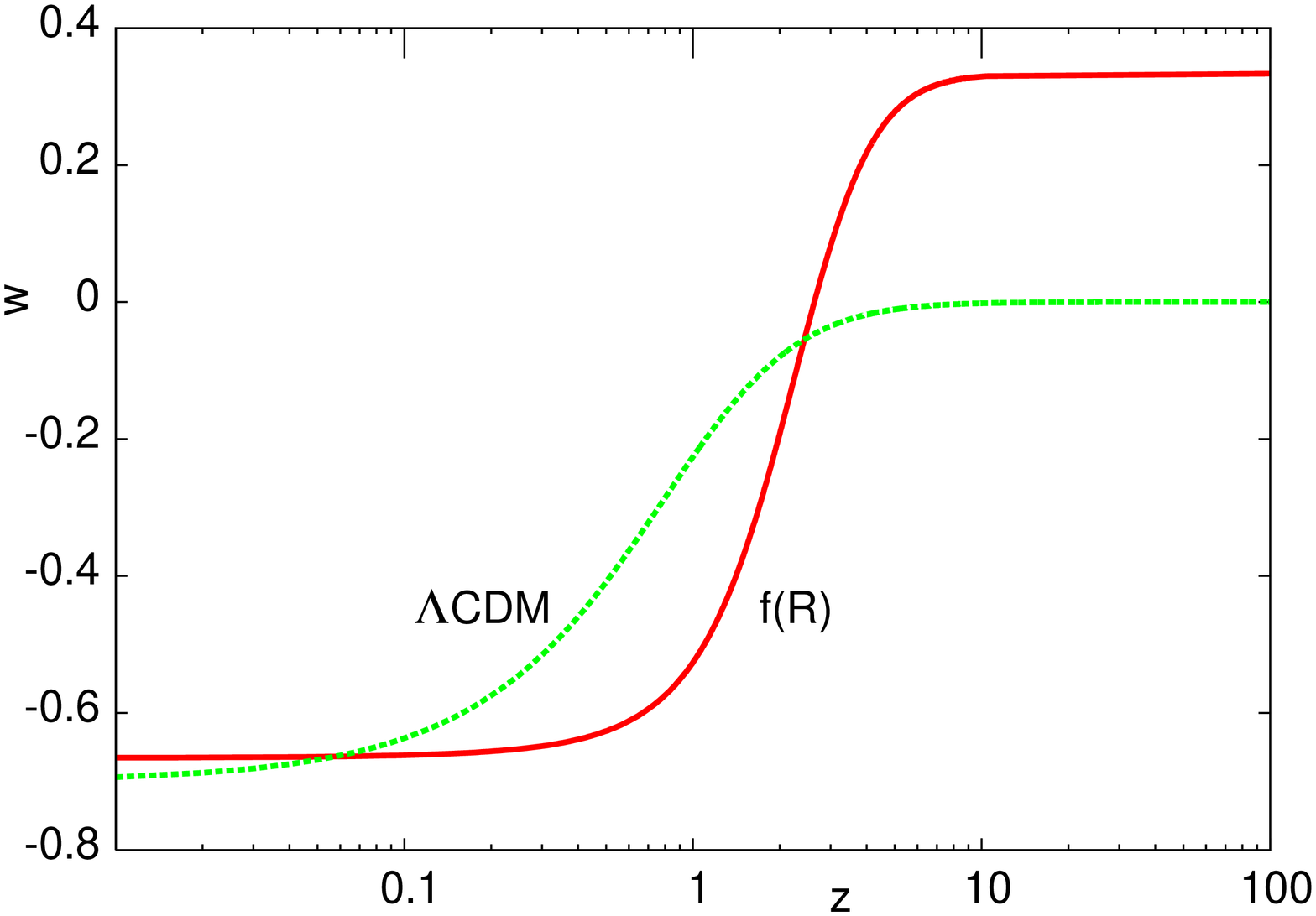}\cr
\end{tabular}    
\caption{\label{hofz}Comparison of the best fit version of $f(R)=R-\mu^4/R$ vs. $\Lambda$CDM, in both cases $\Omega_M=0.3$.  On the left $H(z)/H_0$ and on the right the total effective equation of state $w$ defined in equation (\ref{eos}).}
\end{figure}

The best fit $\chi^2$ values are all rather close to the dividing line to the region where these maximum age cosmologies exist.  In these models, the scalar field runs back from the top of the potential so that $\sigma\rightarrow\infty$, $R\rightarrow 0$ and the effective total equation of state $w\rightarrow 1/3$.  Speaking purely in terms of expansion history, the universe therefore becomes effectively radiation dominated very recently as we go back in time, not because of the energy density of any radiation, but simply because of the modified gravity giving rise to what looks like, from the FRW perspective, an energy density with equation of state $w=1/3$.  This has been noticed by other authors \cite{shinjiamen1,shinjiamen2}.

On the right of figure \ref{hofz} is plotted the effective total equation of state that one would obtain if the same expansion was due to some energy density rather than a different theory of gravity, in other words
\begin{equation}
w=\frac{2}{3}\frac{(1+z)}{H}\frac{dH}{dz}-1
\label{eos}
\end{equation}
and it can be seen that at high redshifts, $w\rightarrow 1/3$ which is the same as pure radiation.

On the left of figure \ref{hofz} we can see that $H(z)/H_0$ is similar in both cases at low redshifts where the supernova and baryon oscillation data is fitted, then the $H(z)$ for the $f(R)$ model dips below, then rises above, the $H(z)$ for $\Lambda$CDM. This means that the luminosity/angular distance integrals between $z=0$ and $z=1000$ are similar.  
\begin{figure}
\begin{center}
\includegraphics[height=10cm,width=14cm]{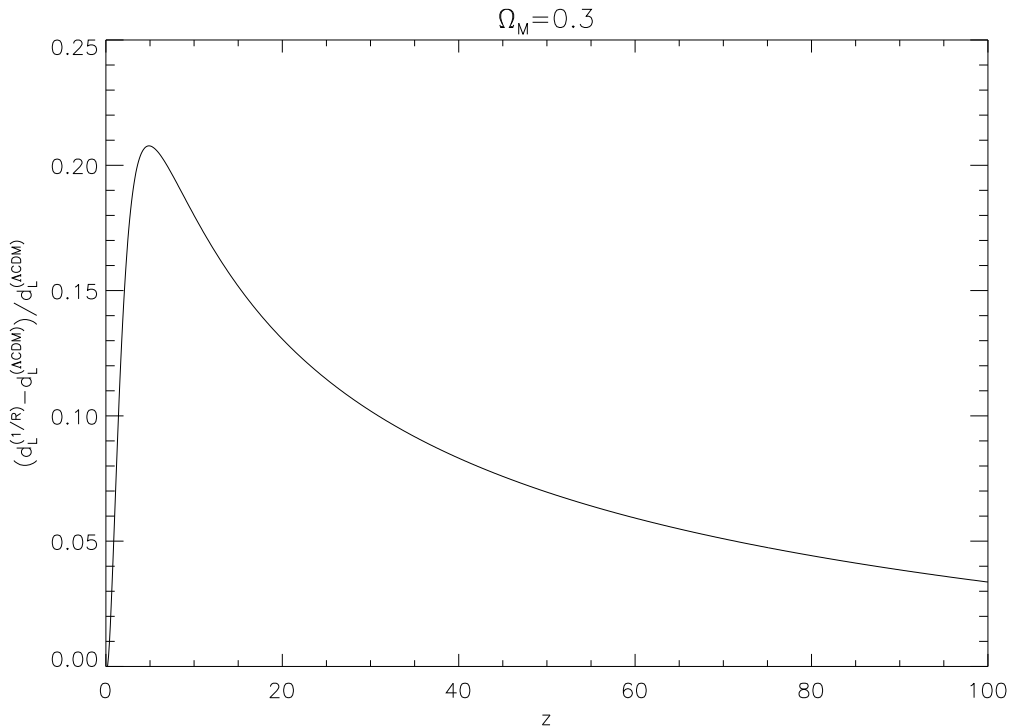} 
\caption{Difference between the luminosity distance as a function of redshift for flat $\Lambda$CDM and the best fit $1/R$ model when $\Omega_M$ is taken to be $0.3$.}
\label{compare}
\end{center}
\end{figure}
This point is made clearer by looking at figure \ref{compare} which shows the fractional difference between the luminosity distance\footnote{and therefore also the angular distance since they are related by a factor of $1+z$} for the $f(R)$ model and the $\Lambda$CDM model.  The difference between the two is rather small at very low and high redshifts but peaks around $z\sim5$.

Better knowledge of the Hubble diagram around redshifts of $z\sim 1-5$ would therefore probably differentiate between the models, but since we only have good data at $z<1.5$ from supernovae and at $z\sim 1100$ from the CMB we are not able to differentiate between them using only these data sets.

Figure \ref{overrres} shows the comparison of the data with the $f(R)=R-\mu^4/R$ model for different values of $\Omega_M$, $\sigma_0$ and $d\sigma/d\tau|_0$.  Plotted are the banded constraints corresponding to the supernova data and to the baryon oscillation data.  A large region of the parameter space is ruled out as it corresponds to regions where $\sigma\rightarrow 0$ too quickly in the past, corresponding to a curvature singularity which would be too recent to accommodate the early universe physics that we know must take place (last scattering surface, nucleosynthesis etc.).  The best fit values of $\chi^2$ for different values of $\Omega_M$ are listed in table \ref{overrtable} and they show that the model works well in obtaining expansion histories rather similar to $\Lambda$CDM.  Addition of the CMB data makes the fit worse but it is still possible to fit the data with a $\chi^2$ per degree of freedom which is less than one.  Also the look back time is consistent with the age of globular clusters \cite{clusterage}.

\begin{table}
\begin{center}
\begin{tabular}{|l|c|c|c|c|c|c|}
\hline
&\multicolumn{3}{|c|}{$R-\mu^4/R$}&\multicolumn{3}{|c|}{$R-\mu^6/R^2$}\\
\cline{2-7}
 \hspace{4.5cm}$\Omega_M=$&0.2 & 0.25 & 0.3&0.2 & 0.25 & 0.3\\
\hline
$\chi^2_{min}$(SNe+flat) & 155 & 155 & 155 & 156 & 156 & 156 \\
$\chi^2_{min}$(SNe+CMB+flat) & 185 & 180 & 177& 173 & 163 & 158 \\
$\chi^2_{min}$(SNe+BAO+flat)  & 169 & 156 & 156 & 175 & 159 & 156 \\
$\chi^2_{min}$(SNe+BAO+CMB+flat)  & 201 & 182 & 179 & 186 & 164 & 159 \\
\hline
\end{tabular}
\end{center}
\caption{\label{overrtable}The best fit values of $\chi^2$ (178 $d.o.f.$ when the supernova data only is included) for the $-\mu^4/R$ and the $-\mu^6/R^2$ modifications of the Einstein-Hilbert Lagrangian.  The vertical columns correspond to different values of $\Omega_M$ while each row corresponds to a different subset of the data.  Flatness is assumed throughout.}

\end{table}

\begin{figure}
\begin{tabular}{cc}
\includegraphics[height=6cm,width=8cm]{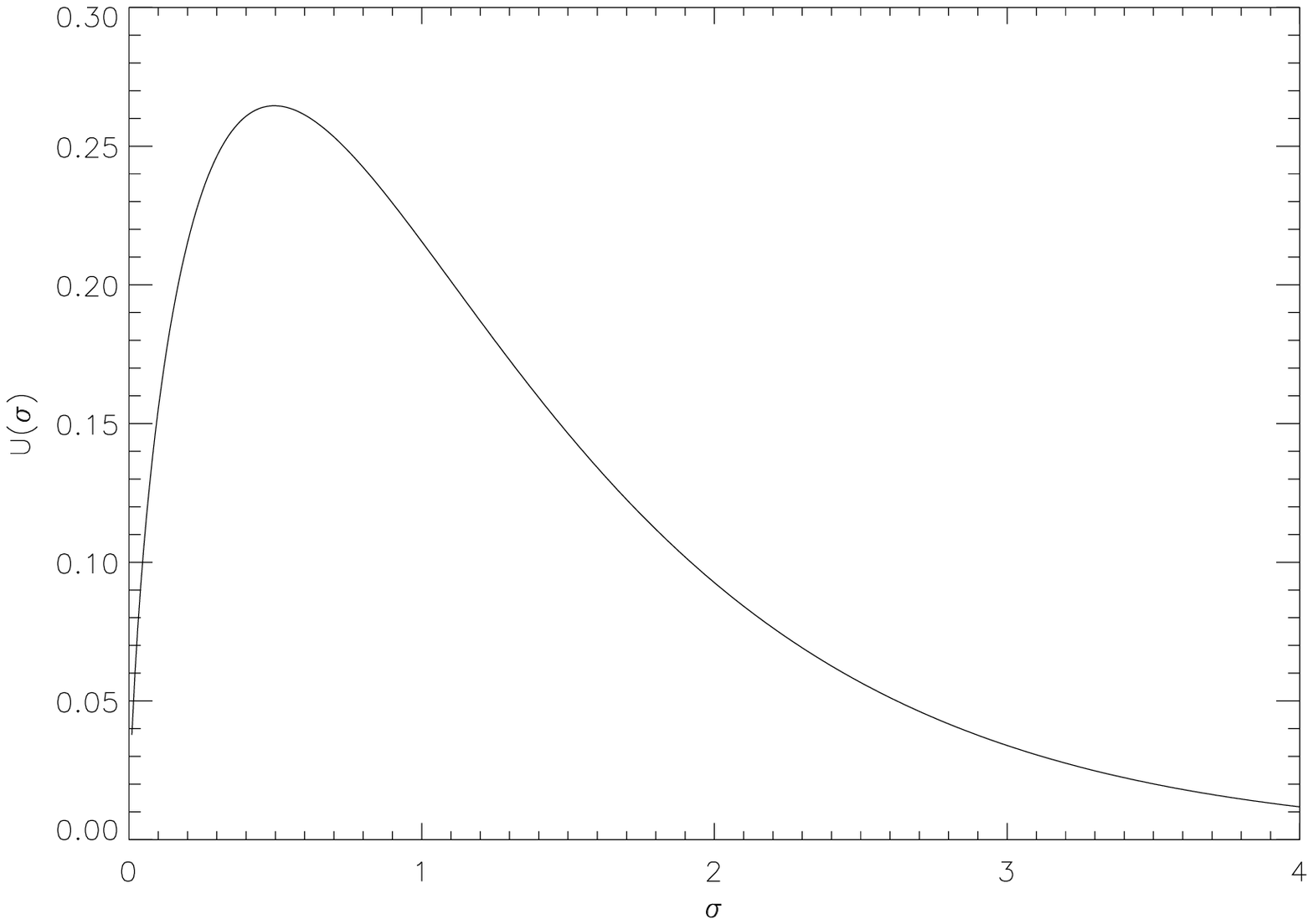}& 
\includegraphics[height=6cm,width=8cm]{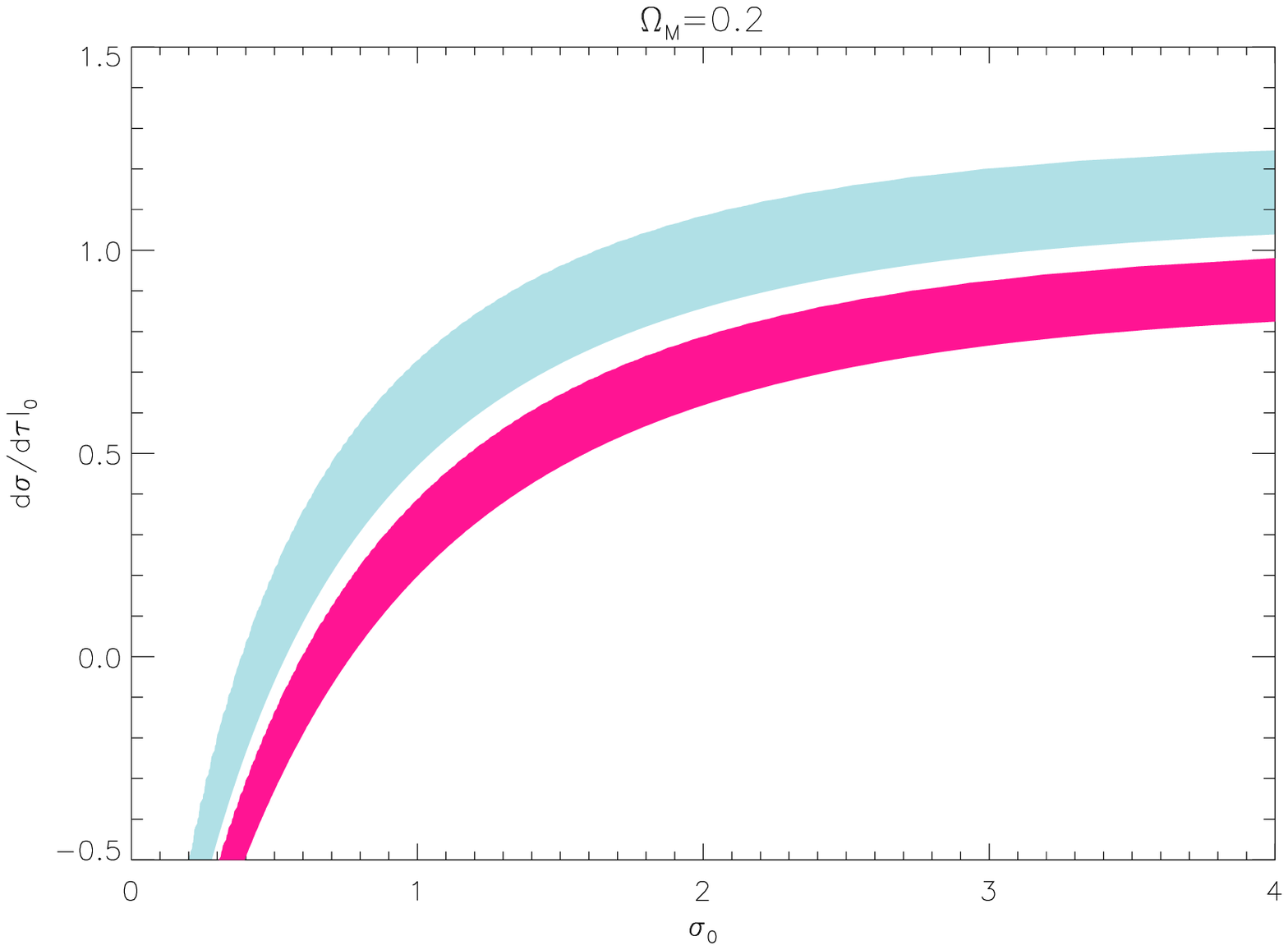}\cr
\includegraphics[height=6cm,width=8cm]{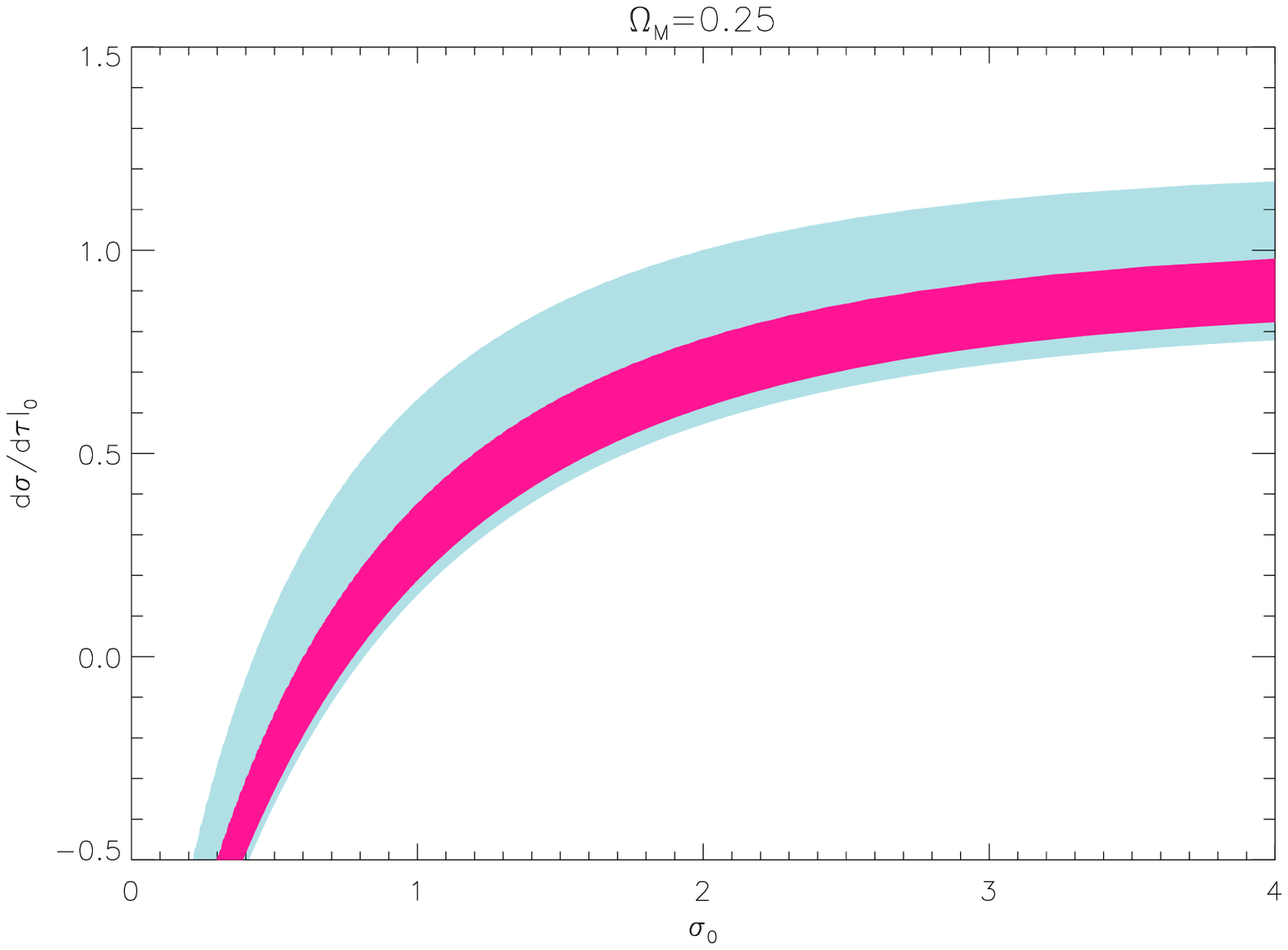}& 
\includegraphics[height=6cm,width=8cm]{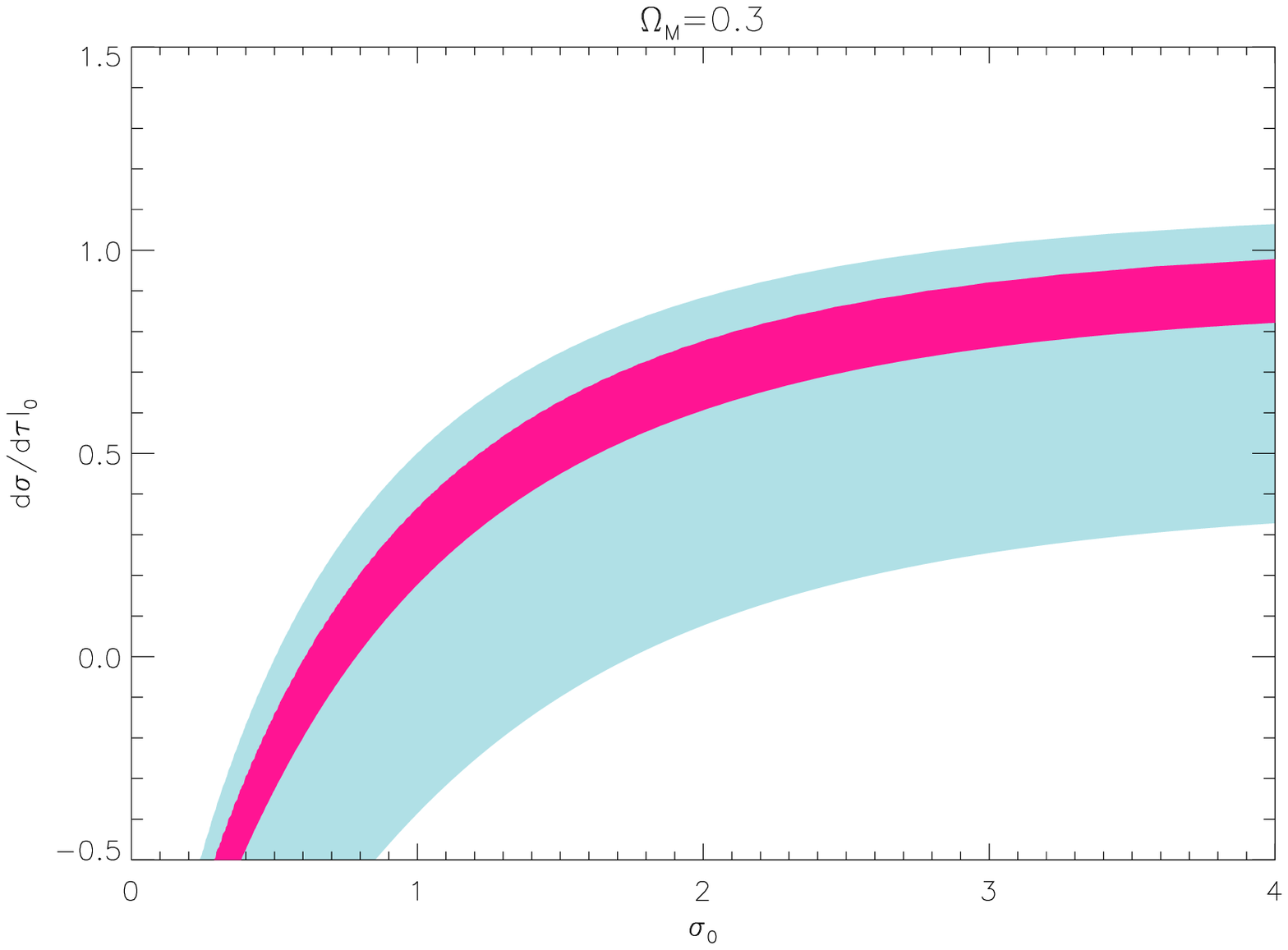}\cr
\end{tabular}    
\caption{\label{overrsqres}Comparison of $f(R)=R-\mu^6/R^2$ with data.  The top left hand panel is the effective potential in the Einstein frame, while the other three plots are the values of $\sigma_0$ and $d\sigma/d\tau|_0$ favoured by the supernova data in red and the BAO data in blue (see caption of figure \ref{lookback} for more details explaining the plot). The three plots correspond to different matter densities - $\Omega_M=0.2,0.25,0.3$.}
\end{figure}

The same analysis has been carried out for the $f(R)=R-\mu^6/R^2$ model of gravity and the results are listed in figure \ref{overrsqres} and the best fit values of $\chi^2$ are listed in table \ref{overrtable}.  The situation is completely analogous to the $-\mu^4/R$ modifications.  Again, the models are able to explain the supernova data.

To summarize this section, universes which fit the data in the $1/R$ model lead to the scale factor $a\propto t^{1/2}$ at high redshift.  Despite this, the universes obtained are old enough to accommodates globular clusters. Also, because the integrals over $1/H(z)$ are dominated at low redshifts, it is possible to obtain expansion histories at low redshifts rather similar to $\Lambda$CDM.

As we mentioned at the beginning of this section, this ability to fit data which is relevant for the $\Lambda$CDM universe is not at all interesting unless the data itself is still relevant.  While this is true for the supernova data, the different behaviour of the scale factor at early times in these inverse $R$ models will change the CMB, and therefore trying to fit the expected position of the first peak that one would expect if the early universe contained only matter and radiation as we have done above would be a mistake.  The simplest way to see how the different expansion in the early universe would affect the CMB is by looking at the thickness of the last scattering surface.

\section{Thickness of the Last Scattering Surface.}

So far it has been shown that in models of modified gravity of the form $1/R$ and $1/R^2$, the expansion of the universe can fit the supernova data.

The age of the universe in these models is perfectly compatible with the age of globular clusters observed.  One therefore cannot use such naive benchmark tests to rule out the model.  The integral of the inverse of $H(z)$ between $z=0$ and $z=1000$ in these models is not too different from $\Lambda$CDM so that features with the same physical size and redshifts as the peaks in the CMB and the galaxy correlation function would posses the same angular size in this universe.  However, this is not encouraging since the physical size baryon oscillations and also the peak in the CMB would be completely different due to different early universe physics.

There are of course still unused weapons in our arsenal of observations, in particular one can think about how the perturbations grow in these models, as has been analysed previously in references \cite{song,bean,li}.  Also one can consider what happens to nucleosynthesis which is in principle rather sensitive to the expansion of the universe during the freeze out of the weak interactions.  

The size of the horizon at the last scattering surface will also be different due to the different expansion which occured at higher redshifts and consequently the time corresponding to that epoch.  This would directly shift the angular position of the acoustic peaks. Furthermore structure formation would be different, leading to very different predictions for the size of the peaks.

However, it is simpler and easier to consider the epoch of recombination and to look at the way that the added expansion at redshifts of order $z\sim1000$ would affect the thickness of the CMB.

The re-combination of protons and electrons which occurs as the universe cools is the process responsible for making the universe transparent.  In normal cosmologies where the universe is still matter dominated at this epoch, the electrons, photons and protons remain rather close to thermal equilibrium throughout the process.  The physics is therefore less dependent upon the expansion rate of the universe but more upon the temperature.  However, the effective radiation density that one obtains at high redshifts for the $f(R)$ models considered here is orders of magnitude larger than the energy density of radiation implied from the temperature of the CMB.  

The density of radiation implied by looking at the $2.7 K$ background radiation is of the order of $\Omega_\gamma\sim 10^{-4}$ so that matter radiation equality occurs at redshifts around a few times $10^4$.   The effective radiation density that fits the $H(z)$ at high redshift correspondent to the best fit $1/R$ model (to the supernova data alone) is $\Omega_{w=1/3}\sim 0.22$.  The expansion rate around $z\sim 1000$ is therefore changed considerably, and recombination takes longer to occur.

The equation which needs to be solved in order to calculate the rate of reionisation is \cite{peebles,turnercmb}
\begin{equation}
\frac{dx_e}{dz}=\frac{1}{H(1+z)}\left[\alpha n_px_e^2-\beta(1-x_e)\exp\left(-\frac{B_1-B_2}{kT}\right)\right]C
\label{boltz}
\end{equation}
where $\alpha$ is the recombination coefficient, $\beta$ is the ionization coefficient and $B_n$ is the binding energy of the $n-$th level of the hydrogen atom.  The factor $C$ is given by
\begin{equation}
C=\frac{1+K\Gamma (1-x_e)}{1+K(\Gamma+\beta)(1-x_e)}
\end{equation}
where $\Gamma$ is the decay rate of the 2s excited state to the ground state via the emission of 2 photons.  The ionisation coefficient $\beta$ is given by
\begin{equation}
\beta=\alpha\left(\frac{2\pi m_e kT}{h^2}\right)^{3/2}\exp\left(-\frac{B_2}{kT}\right)
\end{equation}
and the recombination coefficient $\alpha$ is described by the expression
\begin{equation}
\alpha=\Sigma_{n,l}\frac{(2l+1)8\pi}{c^2}\left(\frac{kT}{2\pi m_e}\right)^{3/2}\exp\left(\frac{B_n}{kT}\right)\int_{B_n/kT}^{\infty}\frac{\sigma_{nl}y^2dy}{\exp(y)-1}
\end{equation}
Solution of equation (\ref{boltz}) gives the ionisation fraction as a function of redshift.  Following \cite{peebles}, we approximate $\alpha$ to be $\alpha\propto T^{-0.5}$ and normalise it so that it gives rise to the correct redshift for ionization as observed by WMAP.  The thickness of the CMB Last Scattering Surface (LSS) can then be found by looking at the probability for finding a photon reaching us from a redshift $z$,  $g(z)=e^{-\tau}d\tau/dz$.  We then define the thickness as being the full width at half the maximum of this function $g(z)$.  In this way, when we consider recombination in the $\Lambda$CDM universe we obtain a thickness for the CMB of $\Delta z=197$ rather close to the WMAP value $\Delta z=195$.

\begin{figure}
\begin{center}
\includegraphics[height=14cm,width=10cm,angle=270]{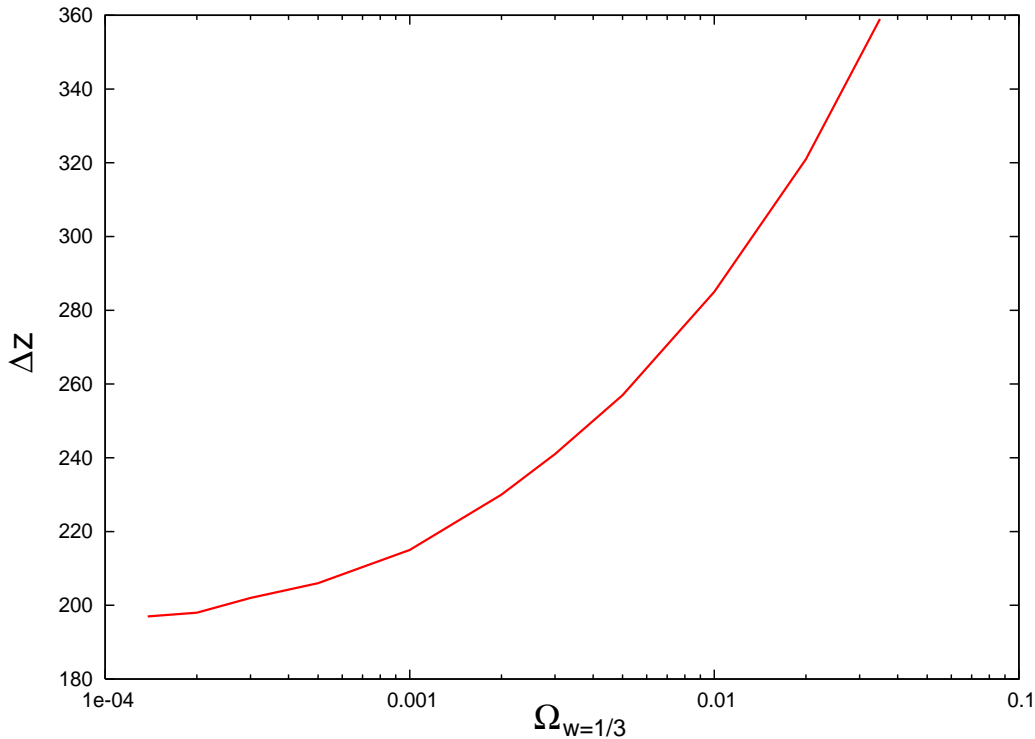} 
\caption{Thickness of the LSS as a function of the effective radiation like energy density.  The red line runs from the value predicted from the CMB in $\Lambda$CDM to the value which best fits the high redshift expansion for the 1/R model which best fits the data.}
\label{thickness}
\end{center}
\end{figure}

Figure $\ref{thickness}$ shows us the thickness of the CMB as a function of the effective radiation density.  The greater the expansion of the universe at last scattering, the more gradual is the era of recombination (gradual with respect to redshift rather than time) and the larger the thickness of the LSS.  The redshift of the LSS also changes, but the fractional change in the thickness is much greater.  The thickness derived from observations by WMAP is $\Delta z=195\pm 2$ \cite{WMAPbasic}, however we find that for the best fit parameters to the supernova data alone, we get a thickness much closer to $\Delta z=530$.
  
The first effect of a shift of the last scattering surface to earlier times would be a different sound horizon size at decoupling, so that the peaks of the CMB would change position \cite{trotta}.  At the same time, the increase in the thickness of the LSS would increase photon diffusion within the surface, blurring out features with a physical size much smaller than the LSS.  Suppression at high multipoles would therefore be brought down to much lower multipoles \cite{hu}.

Constraints upon the total number of relativistic species present which are not chemically coupled to the plasma have been obtained and refined by running codes like CMBFAST which take into account both the effect of the shift in the peak and the supression of higher multipoles (see \cite{trotta} and references therein).  Those studies show that an increase in the number of relativistic degrees of freedom of factor $\Delta N=13.37$ can be ruled out at 2-$\sigma$ by looking at the resulting fit to the CMB data alone.  Since the modified gravity solutions which fit the data found in this paper correspond to an increase in the effective relativistic degrees of around $\Delta N\sim 300$ at high redshift, one can clearly rule them out.

In summary, although the $1/R$ models fit the existing supernova data, they can be ruled out because they will lead to a LSS which is too thick.  The theory therefore is inconsistent with the data it is designed to fit.

\subsection{Finding a solution which fits with the CMB}

We have seen that the solutions which fit the supernova data best lead to 
problems in the early universe due to the extra radiation-like effective 
energy density which occurs at that time.  An important question is 
whether or not it is possible to find solutions to the equations where the 
expansion rate at the redshift corresponding to the last scattering surface 
is comparable to the one predicted in $\Lambda$CDM cosmology, which is in 
good agreement with the observed rate of reionisation.  

We have seen that the thickness of the LSS is increased to $\Delta z\sim 530$ for the best fit to the supernova data alone.  If we try to use the CMB data and the BAO data as angular sizes to fit also,  the thickness of the LSS in the best fit case decreases to $\Delta z\sim 350$.  This is still far to large to make the use of the CMB and BAO angular sizes valid, but it shows that we need to check to see if it is possible to obtain parameters where the thickness of the LSS might be consistent with the CMB data.

\begin{figure}
\begin{tabular}{cc}
\includegraphics[height=10cm,width=15cm,angle=0]{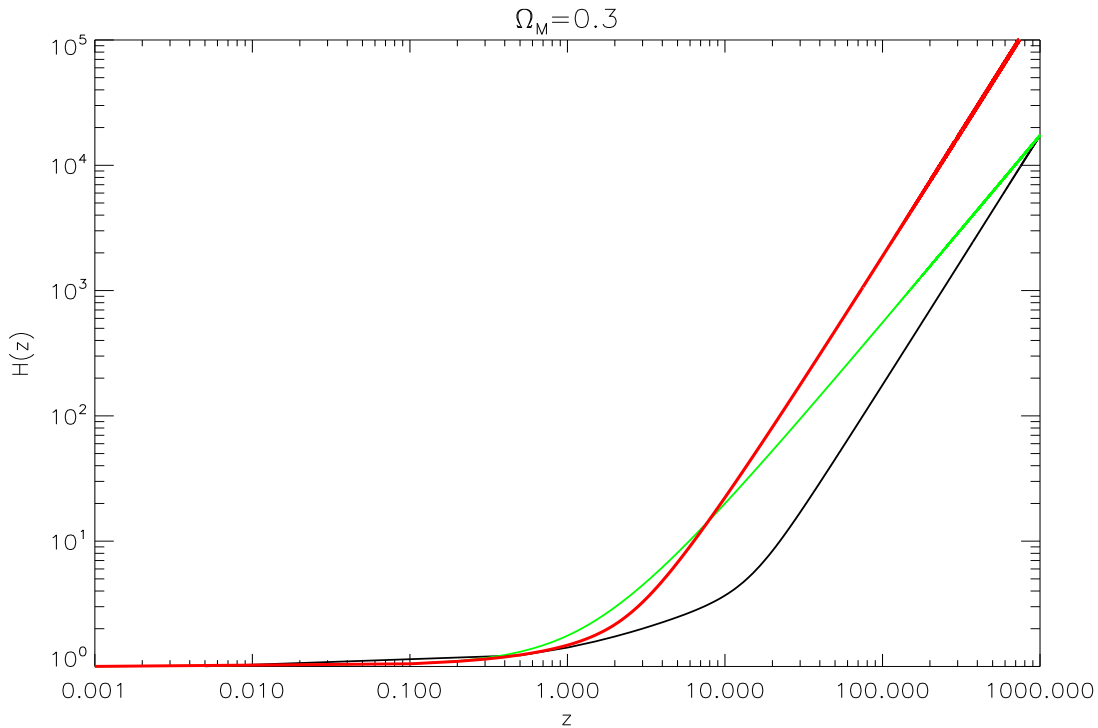}
\end{tabular}    
\caption{\label{hofz2}Comparison of the best fit version of $f(R)=R-\mu^4/R$ (red curve) and $\Lambda$CDM (green curve) vs. a $f(R)=R-\mu^4/R$ solution that is close to the $\Lambda$CDM {\it at the redshift of decoupling} (black curve), in all cases $\Omega_M=0.3$. }
\end{figure}

Figure \ref{hofz2} shows the $H(z)/H_0$ for $\Lambda$CDM, for the best fit $1/R$ model 
mentioned in section \ref{hofz2} and a third cosmology.  This third 
cosmology is a $1/R$ model where the Hubble rate at reionisation is close 
to that experienced in $\Lambda$CDM as can be seen in the figure.  

 The $\chi^2$ fit to the supernova data for these parameters is 184.5, 
which, while much worse than the best fit model described earlier, or the best fit to $\Lambda$CDM, is still close to unity per degree of freedom.

What 
also can be seen is that at lower redshifts, $H(z)$ is much lower than the 
corresponding expansion rate in the $\Lambda$CDM universe.  In fact the 
age of the universe where the expansion rate at the last scattering 
surface fits with observations is $t_0=1.6/H_0$ which corresponds to 25 
Gyrs for $h=$0.7.  In such a universe one would expect to find globular 
clusters with a much lower turn off in the HR diagram then what is 
observed, the current inferred result for the age of the oldest clusters 
being $13.2\pm 1.5$ Gyr \cite{chaboyer}.  The white dwarf cooling sequence 
in globular clusters also points to similar age estimates, making it 
difficult to live with such and old universe \cite{hansen}. 

\section{Summary and Conclusions}

In this paper we have put cosmological constraints on models where the dark energy component of the universe is explained by theories where gravity is modified by the addition of terms to the Einstein Hilbert Lagrangian which contain inverse powers of the Ricci scalar $R$. 

We find that both $1/R$ and $1/R^2$ models, when forced to fit the supernova data, give rise to solutions at high redshifts where the scale factor $a\propto t^{1/2}$.  The universe does not therefore have a matter dominated phase, but rather interpolates between an accelerating phase at low redshifts and a 'radiation-dominated' phase at high redshifts.  The expansion at this earlier phase is driven not by radiation which we know the density of by observing the temperature of the CMB.  Rather this radiation-like expansion is due to the modified Friedman equations in this class of models.  (More complicated models may be able to avoid this feature, see \cite{nojiri}.)

Despite this, we can still find solutions for the Hubble expansion $H(z)$ that are consistent with the recent supernova results.  We have however shown that the theory leads to very different predictions as to what the CMB and therefore the baryon oscillation data should be if it is to fit the luminosity and angular distance data between redshift zero and the last scattering surface.  If we try and force the theory to fit the supernova data while obtaining the correct thickness for the last scattering surface, we are then left with a universe which is much older than the oldest stars observed in our own universe.  We therefore argue that these models are inconsistent with the data.

\ack
We are very happy to thank Tom Dent, Ariel Goobar, Jakob Jonsson, Ignacio Navarro, Jakob Nordin and Michel Tytgat for conversations. MF thanks the Perimeter Institute for Theoretical Physics for their hospitality while some of this work was performed.

\end{document}